\documentclass[letterpaper, 10 pt, conference]{ieeeconf}  

\IEEEoverridecommandlockouts                              

\pdfoutput=1 

\usepackage{graphicx} 
\usepackage{amsmath} 
\usepackage{amssymb}  
\usepackage{color,soul}
\usepackage{multirow}
\usepackage{footnote}
\usepackage[table,xcdraw]{xcolor}
\makesavenoteenv{tabular}
\makesavenoteenv{table}
\usepackage{hyperref}

\newcommand{\toind}{i}
\newcommand{\fromind}{j}

\usepackage{subcaption}

\newcommand{\comm}[1]{}

\newcommand{\Rb}{\mathbb R}

\newcommand{\lb}{\left(}
\newcommand{\rb}{\right)}

\newcommand{\enb}[1]{\lb #1 \rb}

\newcommand{\seq}[1]{ \left\{ #1 \right\} }
\newcommand{\half}[1]{\frac{#1}{2}}

\newcommand{\aver}[1]{\left\langle #1 \right\rangle}

\newlength{\negskiplength}
\newcommand{\Rplus}{\Rb^+}

\newcommand{\bl}{\nu}

\title{\LARGE \bf
	Stability of stochastic finite-size spiking-neuron networks: Comparing mean-field, 1-loop correction and quasi-renewal approximations}

\author{Dmitrii Todorov$^1$, Wilson Truccolo$^{1,2,3}$   
	\thanks{This research was supported by the DARPA program Neural Engineering Systems Design (NESD) (N666001-17-C-4013); National Institute of Neurological Disorders and Stroke (NINDS), grant R01NS079533;  Department of Veterans Affairs, Merit Review Award I01RX000668; The Pablo J. Salame '88 Goldman Sachs endowed Assistant Professorship of Computational Neuroscience. The contents do not represent the views of the U.S. Department of Veterans Affairs or the United States Government.}
	\thanks{$^1$Department of Neuroscience, Brown University, Providence, RI; $^2$Carney Institute for Brain Science, Brown University, Providence, RI; $^3$Center for Neurorestoration and Neurotechnology, Providence, U.S. Department of Veterans Affairs. Email: \{dmitrii\_todorov, wilson\_truccolo\}@brown.edu} 
}

\begin{document}

\maketitle
\thispagestyle{empty}
\pagestyle{empty}

\begin{abstract}
	We examine the stability and qualitative dynamics of stochastic neuronal networks specified as multivariate nonlinear Hawkes processes and related point-process generalized linear models that incorporate both auto- and cross-history effects. In particular, we adapt previous theoretical approximations based on mean field and mean field plus 1-loop correction to incorporate absolute refractory periods and other auto-history effects. Furthermore, we extend previous quasi-renewal approximations to the multivariate case, i.e. neuronal networks. The best sensitivity and specificity performance, in terms of predicting stability and divergence to nonphysiologically high firing rates in the examined simulations, was obtained by a variant of the quasi-renewal approximation.   
\end{abstract}

\section{Introduction}
In recent years, nonlinear Hawkes processes (NHP) implemented as point process
generalized linear models (PPGLMs) have been proven to be a useful tool for
analyzing recordings of neuronal ensemble spiking
activity \cite{truccolo2016point,pillow2008spatio,Truccolo2005,truccolo2010collective}.
These models can capture most of the neuronal spiking patterns and dynamics presented by ODE models, such as the Izhikevich canonical neuron models \cite{weber2017capturing}, while at the same time being easily fit to experimental data via standard optimization tools. 
In addition to their use in revealing encoding properties of neuronal ensembles and in neural decoding for brain-computer interfaces, simulations of these models can also be used in predicting the evolution of neural dynamics in normal or pathological brain states. Furthermore, one can attempt to derive the statistical properties of a given estimated model by computing the corresponding moments and cumulants. However, recent work \cite{gerhard2017stability,chen2018stability} has shown
that often simulation of these models can be unstable, leading to
non-physiologically high firing rates (``runaway excitation''). Thus to make NHPs useful 
for long-term prediction of neuronal activity and simulation studies,
it is important to determine their stability and to understand which model features can lead to non-physiological ranges of dynamics. 
Despite several existing approaches based on stochastic process theory and statistical physics-inspired methods, their performance when assessing the stability and dynamics of data-driven PPGLMs, in particular multivariate models, has not been assessed
yet in detail.




Here, we compare the accuracy of several theoretical approaches for predicting
the occurrence of runaway excitation in multivariate NHPs. Specifically, we consider the following theoretical approaches: mean field
approximation, 1-loop fluctuation expansion \cite{ocker2017linking} 
based on stochastic path integral formulations, and the extension of two quasi-renewal approximations \cite{gerhard2017stability,chen2018stability} to the multivariate case.
These approaches are quite different conceptually, having been introduced in different settings and having limitations in different aspects. We emphasize that we focus on finite-size networks of neurons related to the typical data-driven PPGLMs.


We demonstrate how well these theoretical approaches predict
anomalously stationary high firing rates work when applied to multivariate
PPGLMs whose parameters have been specified in the following ways: randomly generated within a given model class, fitted to nonhuman primate cortex data
and fitted to spiking data subsampled from simulations of a recent model of cortical microcircuits  \cite{potjans2012cell}. 




\section{Methods}
We consider here PPGLMs in the form of multivariate Hawkes processes. These models capture the effects of spiking history of a neuron itself (auto-history) and of cross-history resulting from network interactions. In the Neuroscience context, they relate to other biophysically inspired models such as the spike response model \cite{gerstner2014neuronal} and to nonlinear LIF neuron models with adaptation (sometimes referred to as generalized integrate and fire model, GIF \cite{pozzorini2015automated}), but without explicit reset mechanism for the membrane potential.

\renewcommand{\Rplus}{\Rb_{\geq 0} }
\newcommand{\Rplusbar}{\overline{\Rb}_{\geq 0} }

We define the nonlinear Hawkes process in the usual way (e.g.
\cite{bremaud1996stability}), except for enforcing an absolute refractory
period, which is a well known biophysical property of biological neurons. We
discuss it's implications below. For a given number neurons or dimension $d
\geq 1$, a multivariate Hawkes process $N(t)$ is defined by the
conditional intensity functions
\begin{gather}
  \lambda_{\toind}(t \vert N( [-\infty,t))^d) := \nonumber \\ 
  \lim_{\Delta t\to 0} \frac{1}{\Delta t} P\enb{ \left. N_{\toind}(t+\Delta t) - N_{\toind}(t) = 1 \right| N( [-\infty,t))^d  }  
  \nonumber
  \\
  =
    \phi\enb{ \sum_{\fromind}^d (\eta_{\fromind \to \toind} * dN_{\fromind}) (t) + \bl_{\toind} }, 
    \label{eq:Hdef}
\end{gather}
where $0\leq \toind, \fromind \leq d, \, t\in \Rb$, $\eta_{\fromind\to \toind} : \Rb \to \Rb \cup \seq{-\infty}$ is a continuous function (on the set where it 
takes finite values) and is finitely supported 
$\mathrm{supp}\ \eta_{\fromind\to \toind} \subset (0,\varkappa) $, for $\varkappa > 0$; in other words, 
the $\{\eta_{\fromind\to \toind}\}$ correspond to causal temporal filters that capture spiking history effects.
$dN_{\toind}(t)$ is the increment process, i.e. the spike train, for the ith neuron;
$\bl \in \Rb^d$;  $\phi$ does not mix coordinates; and `$*$' denotes a
convolution.  We emphasize several choices we made in our study and other
related issues. 

First, typically the nonlinearity $\phi(x) $ is taken to be one of the following functions: $a\max(0,x),a\max(0,x)^2,a e^x$ for fixed $a>0$, so that the the requirement that the conditional intensity function (instantaneous spiking rate) be nonnegative is ensured, especially when considering inhibitory effects. Here, we will consider only the case $\phi(x) = e^x$, since it is commonly used in PPGLMs.

Second, we enforce the absolute refractory period by requiring auto-history filters $\eta_{\toind\to \toind}(t)=
-\infty$ for $0\leq t \leq \tau_{r}$, where $\tau_{r} $ is the absolute
refractory period. This is the only reason we allow $\eta_{\toind\to \toind}$ to take negative infinity values. Note that this way we have a uniform bound on the
value of the conditional intensity function and also on the number of event in
a time interval of fixed length. Using a common value $\tau_{r} =
2$ ms, the upper bound on the intensity function is $500$ spikes/s. Thus, all
models we consider here are strictly stable. Nevertheless, we henceforth refer
to \emph{divergent} those cases of non-physiologically high firing rates, i.e.
much higher firing rates than typically observed in recorded neurons.
Here, we define divergence to non-physiological rates when the rate
exceeds $450$ spikes/s. 

Third, in contrast to standard point process theory
\cite{daley2007introduction}, where the conditional intensity function is
strictly positive, here we allow the intensity to approach zero since we
enforce an absolute refractory period as discussed above.

Finally, we also note that the issue of stationarity of NHP processes
warrants further examination. We note that the classical results from
\cite{bremaud1996stability} are not applicable here because, among other
reasons, their requirement of the conditional intensity being strictly positive
is no longer satisfied because of the enforced absolute refractory. Rigorous
proofs of stationarity properties are outside the scope of this paper.
Nevertheless, we note that our simulations show that the processes we consider
here have one or several stationary states, i.e. after some simulation time the
processes converge to a steady-state of roughly constant mean rates and
fluctuation size. Therefore, we conjecture that a process \eqref{eq:Hdef}
with refractory period may have finitely many ergodic components (and of course
maybe stationary within each of them). Since we are not planning to prove
stationarity and ergodicity, we will assume their presence. In addition,
ergodicity is a subtler issue for point processes \cite{baccelli2013elements}.
Here, we use the term ``ergodicity'' in its less formal sense commonly stated
as ``averages over realization equal averages over time.'' 
%
%
%
In the next sections, we introduce the theoretical approximations we examined here to determine the stability of multivariate NHPs.

\subsection{Mean field (MF)}

Mean field approximations usually involve some sort of weak coupling assumption, i.e. the strength of connections should decrease as the number of neurons increases. However since we aim at applying these approximations to data-driven PPGLMs, we are constrained to deal with (finite-size) networks whose coupling strengths reflect statistical dependencies as seen in the data. Therefore, since here we assume stationarity, we can work with a ``temporal mean field'' approximation \cite{gerstner2014neuronal}. It means that instead of averaging activity over neurons, we average the activity of every given neuron over all realizations of its past activity. Stationarity then means that such average is constant over time.

More formally, we define a new NHP on $\Rb^d$ using a modification of the initial definition \eqref{eq:Hdef}
  \begin{gather*}
    \lambda_{\toind}^{MF}(t \vert N( [-\infty,t)^d )) := \\
   \phi\enb{ \sum_{\fromind}^d \enb{ \int \eta_{\fromind \to \toind}(s) ds} \aver{ dN_{\fromind}(t)}  + \bl_{\toind} },
  \end{gather*}
where $\aver{dN_{\fromind}(t)}$ denotes the temporal average of the spike train up to time $t$, which is constant due to the stationarity assumption and we denote it by $\bar{r}_j$.  

Ultimately, we arrive at the following self-consistent equation \cite{ocker2017linking}, which can be interpreted as an input-output map $F_{MF}$ or a discrete-time dynamical system on $\Rplus^d$
\begin{gather}
    F_{MF}(\bar{r}) := \phi\enb{ \enb{\int \tilde{\eta} } \cdot \bar{r} + \bl },
    \label{eq:FMF}
  \end{gather}
where $\int \tilde{\eta} $ is a matrix of integrals with filters that have $-\infty$ values (refractory period) replaced by $0$ values, and $\cdot$ means matrix multiplication. Removal of $-\infty$ values is necessary, otherwise the integrals do not converge. To justify it, one can multiply each of the integrals by $ (\varkappa - \tau_{r} )  / \varkappa $ to account for removing the average activity from the mean. However, we work with the case where $\varkappa$ is much larger than $\tau_{r}$ so that the multiplier above is very close to $1$ and has little impact.

\subsection{MF + 1-loop correction (MF+1L)} 

In some sense, the above ``temporal'' mean field approximation requires all couplings between moments of different orders to be neglected. Alternatively, one can include some of these couplings in higher-order approximations in different ways. We consider here the ``1-loop correction,'' a method from stochastic field theory extended by \cite{ocker2017linking} to multivariate nonlinear Hawkes processes. 


Basically, the idea is to represent a Hawkes process with a discrete-time stochastic distributed delay differential equation with nonlinear noise, and then use standard approaches used to analyze nonlinear Langevin equations. We refer the reader to \cite{ocker2017linking} for details. In addition, we modify the 1-loop correction formula derived in \cite{ocker2017linking} in order to handle absolute refractory periods. 

First, an approximation of the response function $R(t,t')$, which indicates how stationary average rates of the process would change to a perturbation to the stochastic drive at time $t'<t$, is derived as 
  \begin{gather*}
      R_{\toind\fromind}(t,t') := 
      \lim_{\substack{ \Delta t \to 0, \\ h\to0 }} 
      (   \aver{ N_{\toind}(t) - N_{\toind}(t+ \Delta t) }  - 
      \\
    - \aver{  N_{\toind,\fromind,t',h}(t ) + N_{\toind,\fromind,t',h}(t+ \Delta t )  }  ) /
    ( h \Delta t ) , 
  \end{gather*}
 where $N_{\toind,\fromind,t',h} $ is $\toind$-th coordinate of the modified NHP $N$ --  
 the constant (offset) drive $\nu_{\fromind}(t) \equiv \nu_{\fromind}$ was 
 replaced by a time-dependent one $\nu_{\fromind} + \delta(t-t') h $.

In the context of Hawkes processes it is in fact better to think about the response function as a response to adding a spike at a past time $t'<t$, since in this model all neuron-to-neuron
interactions happen only via action potentials (i.e. the response to a drive change can only happen due to the occurrence of a spike in this or other neuron). This means, in particular, that $R_{\toind\toind}(t,t') = 0$ for $t-t' < \tau_{r} $, as a neuron cannot respond to input spikes arriving during the absolute refractory period.

In the standard Langevin approach, the first approximation $\bar{\Delta}$ to
the response function $R$, sometimes called ``linear response'', or
``tree-level'' response \cite{ocker2017linking}, can be obtained by solving the following equation:
\begin{gather*}
 \bar{\Delta}_{\toind\fromind}(t-t')  = \phi^{(1)}(\bar{r}_{\toind} )  
    \sum_k (\eta_{k \to i} * \bar{\Delta}_{kj} )(t-t')
    + \delta(t-t')\delta_{\toind\fromind},
 \end{gather*}
 where $\phi^{(1)}$ is the first-derivative of the nonlinearity.
  
If $\eta$ has finite values, then the equation is not difficult to solve by replacing  convolutions with the multiplications in the Fourier domain. However, to avoid problems with $-\infty$ values related to refractory period, we use the following equation instead:

\begin{gather*}
    \tilde\Delta_{\toind\fromind}(t-t')  = \phi^{(1)}(\bar{r}_{\toind} )  
    \sum_k (\tilde\eta_{k \to i} * \tilde\Delta_{kj} )(t-t')
    + \delta(t-t')\delta_{\toind\fromind}.
\end{gather*}



 Now the 1-loop correction $\tilde{r}_1$ can be defined as 
 \begin{gather}
   \tilde{r}_1 \equiv r_1(t) =  \half{1} \int_{0}^t   {\tilde{\Delta} }(t-t_1) \cdot
   \label{eq:1L}
   \\
   \cdot \phi^{(2)} \enb{ \nu+\enb{\int\tilde{\eta} }\bar{r}} 
   \int_{0}^t \enb{  (\tilde{\eta} * {\tilde{\Delta} })(t_1-t_2)}^2 \bar{r}\   dt_2\,  dt_1,
   \nonumber
 \end{gather}
 with the MF+1-L approximation corresponding to 
  $ \bar{r}^{(1L)} := \bar{r} + \tilde{r}_1. $
%
Above we use vector notation: $\phi^{(2)}(\bar{r}) $ is a diagonal matrix with
diagonal entries being second derivative of $\phi$ evaluated at different coordinates of $\bar{r}$; the matrix-valued 
function $\tilde{\eta}*\tilde{\Delta}$ is the
result of coordinate-wise convolution.

 Replacing $\eta$ by $\tilde{\eta}$ to address infinities resulting from
absolute refractory periods corresponds to considering a process without
refractory period at all (but which still takes into account all other auto-history effects encoded in the remaining parts of the auto-history filters). 
We will show that in some cases such approach still works despite that.

Note that the formulas in this section do not formally require $\bar{r}$ to be a mean-field map fixed point. Therefore, we will also consider a linear response $\Delta^{(\bar{q})}$ for any stationary rate approximation $\bar{q}$, as well as 1-loop correction around $\bar{q}$.



 \subsection{Event-based moment expansion (EME1)}

    
In the above mean field approximation, the average brackets were placed inside
the nonlinearity. Alternatively, this average can be placed outside, leading to
the EME1 approximation \cite{naud2012coding} if $\phi = \exp$. The resulting
expectation $\aver{ \exp( dN * \eta )(t) }_{N_{[-\infty,t)}}$ taken with respect to all
    possible spike histories corresponds to a moment generating functional,
    evaluated at source $=\eta$. This alternative approximation
    \cite{naud2012coding,van1992stochastic} leads to a temporal mean-field type
    formula 
  \begin{gather}
    F_{EME1}(\bar{r}) := \phi\enb{ \enb{\int_0^{\infty} \exp(\eta) - 1} \cdot \bar{r} + \bl }.
    \label{eq:FEME1}
  \end{gather}

\subsection{Quasi-Renewal (QR)}

In the QR approximation
\cite{naud2012coding,schwalger2017,gerhard2017stability}, we start by
considering auto-history effects of the most recent
spike and by averaging over all previous spiking histories sharing this most
previous spike. Here, cross-history effects (neuronal interations) are approximated by averaging over cross-spiking histories as done in \cite{schwalger2017} for their GIF-network models. Specifically,
\[
		\begin{aligned}
		\lambda_i(t\vert N[-\infty,t)^d ) \approx & \exp \enb{ \nu_i + \eta_{i \to i}(t_i')} \times \\
		&\aver{\exp\enb{(\eta_{i \to i} * dN_i) (t_i')}}_{N_{[-\infty,t_i')}} \\
		&\times \exp \enb{\sum_{j\neq i}^d \aver{(\eta_{j \to i} * dN_j) (t)} }, 
		\end{aligned}
\]		
\comm{
  \begin{gather*}
  \lambda_{\toind}(t\vert N( [-\infty,t)^d )) \approx
    \exp \left( \bl_{\toind} + \eta_{\toind \to \toind}(t'_{\toind})  +  
    \vphantom{ \aver{(\eta_{\toind \to \toind} * dN_{\toind}) (t_{\toind}')}_{N_{[-\infty,t_{\toind}')}}   
      + \sum_{\fromind\neq \toind}^d \aver{(\eta_{\fromind \to \toind} * dN_{\fromind}) (t'_{\fromind})} }
    \right.
    \\
    +
    \left. \aver{(\eta_{\toind \to \toind} * dN_{\toind}) (t_{\toind}')}_{N_{[-\infty,t_{\toind}')}}   
      +
      \sum_{\fromind\neq \toind}^d \aver{(\eta_{\fromind \to \toind} * dN_{\fromind}) (t)} \right)  
   \end{gather*}
}
where $t'_{\toind}$ is the time of the most recent spike of the $\toind$'th neuron. As stated above, we remind that the temporal filters $\{\eta_{\fromind\to \toind}\}$ are time causal for all $i, j$. 

  %


Next, the expectation involving the auto-history term is approximated by the
1st order term of the expansion of the corresponding moment generating
functional expansion, similarly as done for EME1 (see also \cite{gerhard2017stability} for details). 

The resulting QR approximation for the conditional intensity function is then
\begin{gather}
  \lambda_{\toind}^{QR}(s,A) :=  
  \exp \left( \bl_{\toind} + \eta_{\toind\to \toind}(s)+ 
  \vphantom{ A_i \int_{s}^{\infty} (e^{\eta_{\toind\to \toind}(z)}-1) \, dz   + \sum_{\toind\neq \fromind}^d A \int_{0}^{\infty} {\eta_{\fromind\to \toind} (z) } }
  \right.
  \label{eq:lambda0}
  \\
  \left.
 + A_i \int_{s}^{\infty} (e^{\eta_{\toind\to \toind}(z)}-1) \, dz   
  + \sum_{\fromind\neq \toind}^d A_{\fromind} \int_{0}^{\infty} {\eta_{\fromind\to \toind} (z) }  \, dz   \right), 
  \nonumber
\end{gather}
where $A$ is a vector whose components are given by the average mean firing rate of the corresponding neurons (recall that we consider a stationary situation), and $s = t - t_{\toind}$ is the time elapsed since the most recent spike. 


The above expression can be simplified by removing the dependence on $s$. For the auto-history 
contribution one can proceed as in standard renewal theory: compute the steady-state survival function and corresponding ISI probability density based on the approximation; the inverse of the corresponding mean ISI gives the predicted firing rate (Eq. 8 in \cite{gerhard2017stability}), which corresponds, under stationarity assumption, to a self-consistent integral equation or input-output map $F_{QR}:\Rb^d\to \Rb^d$ 

\begin{gather}
  F_{QR}(A)_{\toind} :=  
  \enb{ \int_0^\infty e^{-\int_0^\tau \lambda_{\toind}^{QR}(s,A) ds} d\tau }^{-1}.
  \label{eq:FQR}
\end{gather}

We note that in \eqref{eq:lambda0}, the integral inside the exponential function has the property
that positive values of $\eta$ contribute much more than the negative values.
Therefore, the QR approximation weights more the contribution of the positive
phases of the temporal filter when computing the averaged dynamics prior to the
most recent spike.

\subsection{Quasi-Renewal-MF (QRMF)}

Finally, the QRMF approximation is similar to the QR approximation, except that the expectation over the spiking histories prior to the most recent spike is taken inside the nonlinearity $\phi$ rather than outside \cite{schwalger2017,chen2018stability}. In other words, it is akin to a temporal mean field version of the QR approximation. This approximation was first introduced in \cite{chen2018stability} for the single-neuron case. Here, we extend it to multiple neurons using the same approach as above -- adding mean field cross-interactions. Modifying \eqref{eq:lambda0} accordingly, we obtain
\begin{gather*}
  \lambda_{\toind}^{QRMF}(s,A) :=  
  \exp \left( 
  \vphantom{ 
  A \int_{s}^{\infty} \eta_{\toind\to \toind}(z)  dz  + 
  \sum_{\toind\neq \fromind} A \int_{0}^{\infty} {\eta_{\fromind\to \toind} (z) }  dz   }
  \bl_{\toind} + \eta_{\toind\to \toind}(s)+ \right.
 \\
 \left.
 +
   A_i \int_{s}^{\infty} \eta_{\toind\to \toind}(z)  dz  + 
  \sum_{\fromind \neq \toind} A_{\fromind} \int_{0}^{\infty} {\eta_{\fromind\to \toind} (z) }  dz   \right). 
  \label{eq:lambda0MF}
\end{gather*}

The corresponding map is
\begin{gather}
  F_{QRMF}(A)_{\toind} :=  
  \enb{ \int_0^\infty e^{-\int_0^\tau \lambda_{\toind}^{QRMF}(s,A) ds} d\tau }^{-1}   .
  \label{eq:FQRMF}
\end{gather}

\subsection{Implementation}

We developed Python/C software that computes all the approximations in a systematic way. It also performs simulations and fitting of PPGLMs to spike data from neuronal ensembles. We hope to make the codes available in a future more extend publication. Refractory periods were implemented as
a sufficiently large negative value for the temporal filters for time lags
$(0,2]$ ms. Accordingly, all the dynamical maps described above were capped to
have maximum value of $500$ spikes/s for every neuron.  The formula for 1-loop
correction was constrained to values in $[0, 500]$ spikes/s to ensure both
refractory period and non-negative firing rates.

We also constrained the temporal filters for all neurons in a given network to
be constructed from a common set of $K$ basis functions. One of the common
choice is raised cosines basis \cite{pillow2008spatio,truccolo2010collective},
which allows the efficient fitting of PPGLMs to spike train data. Here,
inspired by \cite{duarte2016stability}, we use a different class of basis
functions, specifically Erlang basis functions. In contrast to
\cite{duarte2016stability}, we introduce a time shift related to the absolute
refractory period, so that each $k$-th basis corresponds to 
\begin{gather*}
\beta_k(t) = H(t-s_k) (t-s_k)^{m_k} e^{-(t-s_k)\alpha_k},
\end{gather*}
where $s_k$ is a time shift, $H(\cdot)$ is the Heaviside function, and $m_k$
and $\alpha_k$ are parameters related to smoothness/rise speed and decay speed, respectively. (We fixed $m_k$ and $\alpha_k$ to values that resulted in basis functions similar to the above mentioned raised-cosine functions).
Each filter $\eta_{\fromind\to \toind}$ is then constructed as the sum of $K$ fixed basis functions 
\begin{gather*}
  \eta_{\fromind\to \toind} (t) = \sum_{k=1}^{K} J_{\toind,\fromind,k} \beta_k(t),
\end{gather*}
with coupling parameters $\{J_{\toind,\fromind,k}\}$. Therefore in our formulation, a $d-$dimensional nonlinear Hawkes process $N$ contains $d$ offset parameters $\bl$ and $K$ basis functions per temporal filter, being characterized by a total of $d + K\times d^2$ parameters.



 

%
%
%
%
%

\section{Results}
After extending existing approximation approaches to determine
NHP stability and dynamics of neuronal networks (Methods), we were able to study the stationary
behavior in the context of {discrete-time dynamical systems} or maps given by Eqs. \eqref{eq:FMF} - \eqref{eq:FQRMF}. We compared the
ability of different approaches to {predict divergence} to non-physiological
firing rates, defined here as rates greater than 450 spikes/s. 
We also examined how different model features might contribute to
divergence or to the performance of the different approximation approaches.

We computed fixed points by forward iteration of the input-output maps. 
We used 1-2,000 initial conditions, 
with two of those deterministically selected (setting one of the 
coordinates to $450$ or $500$ spikes/s and the other to zero); all remaining conditions were randomly sampled (uniformly) from a $[0,500]^d$ volume (unit in spikes/s).

Next, the stability of revealed fixed points was then determined by examining the largest eigenvalue modulus of the derivative of the corresponding map at the fixed point.


As indicated earlier, one can make a 1-loop correction around any vector of firing rates, not necessarily the ones corresponding to the fixed points of the MF map. Thus, we also considered this possibility, henceforth referred to as ``(approx)+1L'', in the 1D (single-neuron) case to compute the 1-loop correction around MF, EME1, QR, QRMF using formula \eqref{eq:1L}. 
Following \cite{ocker2017linking}, the stability of the above approximations with the 1-Loop correction is given by looking at the eigenvalues of the 
stability matrix, corrected by the 1-loop contribution, computed similarly as in \eqref{eq:1L}. Given an approximation of a stationary rate $\bar{q}$, the 1-loop correction to a 
(mean field-,QR-,QRMF-, or EME1-derived) stability matrix $\Psi$ is

\begin{gather}
\Gamma_1=\frac{1}{2} \phi^{(2)} \left(\nu+\left(\int\tilde{\eta} \right) \bar{q} \right)
\int_{0}^t     
\int_{0}^t 
\left(  (\tilde{\eta} * {\tilde{\Delta} }^{(\bar{q})} )(t_1-t_2)\right)^2 \nonumber \\
\phi^{(1)}\left(\nu+\left(\int\tilde{\eta} \right) \bar{q} \right) \tilde{\eta} \   dt_2 \, dt_1,
\label{eq:FQRMF}
\end{gather}
where $\phi^{(2)}$ is the second-derivative of the nonlinearity. The one-loop corrected stability matrix is then

\begin{gather*}
\Psi_1 = \Psi + \Gamma_1.
\end{gather*}

After computing all the approximations as described in the Methods section, we have a predicted number (possibly zero) of stationary mean firing rate values and corresponding predictions for stability (divergence to nonphysiological rates) according to each one of them.

\subsection{Main steps and simulations}
To determined stability (or divergence to nonphysiological rates) based on the dynamical map, we use the following convention:  if any of the found stable fixed points has rate above $450$ spikes/s in at least one of the coordinates, we say that the approximation predicts divergence.
For MF+1L approximation, we judge the model divergent 
if any of two conditions is satisfied: 1) one of the corrected
MF rates is stable in 1-loop correction sense and has rate above $450$ spikes/s in at least one of the coordinates or 2) there are no corrected MF rates which are stable in the 1-loop sense with rates below $450$ spikes/s.

We checked the validity of predicted divergence according to the theoretical approximations by simulating the corresponding NHP models. To detect divergence in simulations we ran $20$ realizations for $80$ or $200$ seconds for dimensions $D \leq 2$ or $D = 100$, respectively. A simulation was considered divergent if either its average spiking rate was above $450$ spikes per second for some of the coordinates, or at some point during the simulation more than $450$ spikes were generated within a $1$-second window. 


\subsection{Mutistability}



\begin{figure}
      \small
    \centering
      \includegraphics[width=0.47\textwidth]{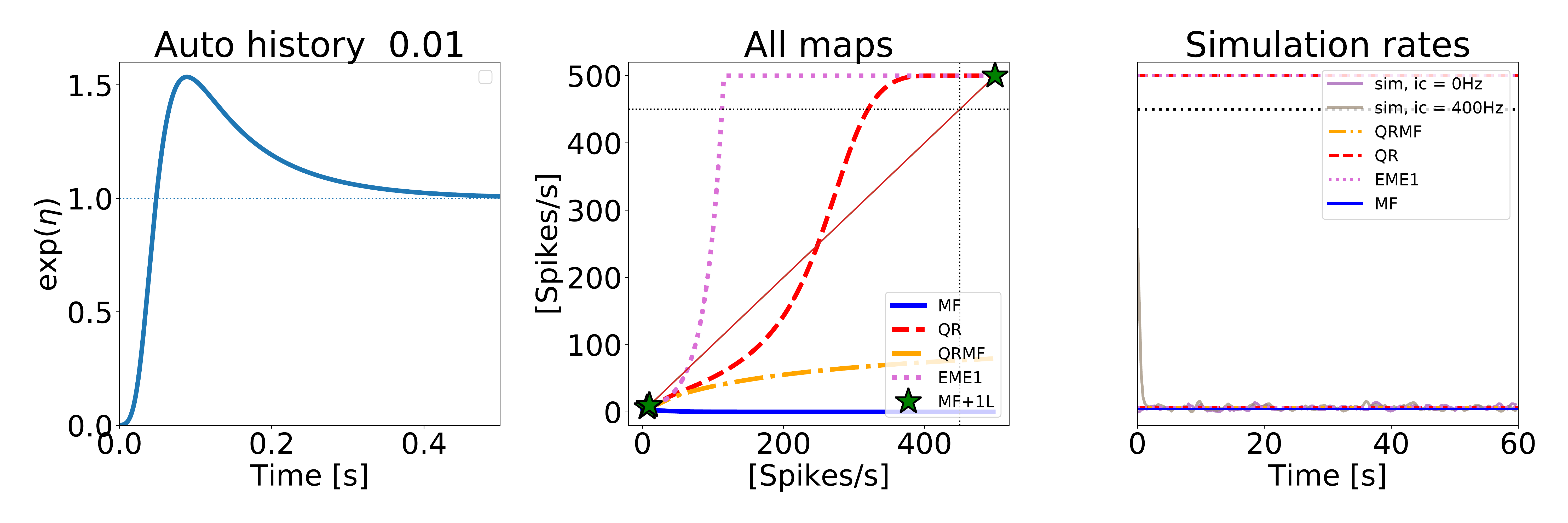}
      \caption{ 1D case: MF, EME1, QR, QRMF maps. EME1 and QR wrongly predicts multistability and divergence (rate $> 450$ spikes/s), while MF and QRMF do not. The auto-history filter was given by the sum of two exponentials and an absolute refractory period. The term ``ic'' denotes initial conditions.}
     \label{fig:QRworse}
    \end{figure}

\begin{figure}
    \centering
    \vskip3pt
    \includegraphics[width=0.45\textwidth]{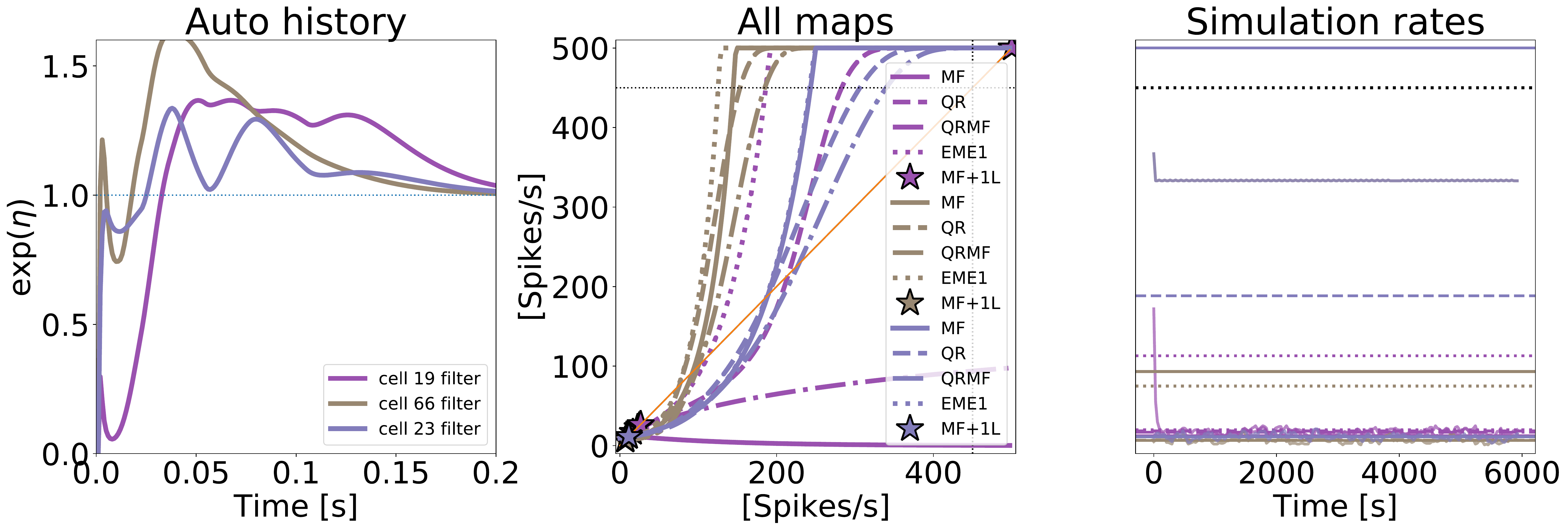}
    \caption{ Three 1D examples of data-driven NHP models fitted to cortical
      neurons (from monkey recordings) that pass time-rescaling goodness-of-fit
      tests \cite{gerhard2017stability}, but whose simulations might diverge.
      %
  }

    \label{fig:fit}
  \end{figure}

\begin{figure}[h]
    \centering
    \includegraphics[width=0.45\textwidth]{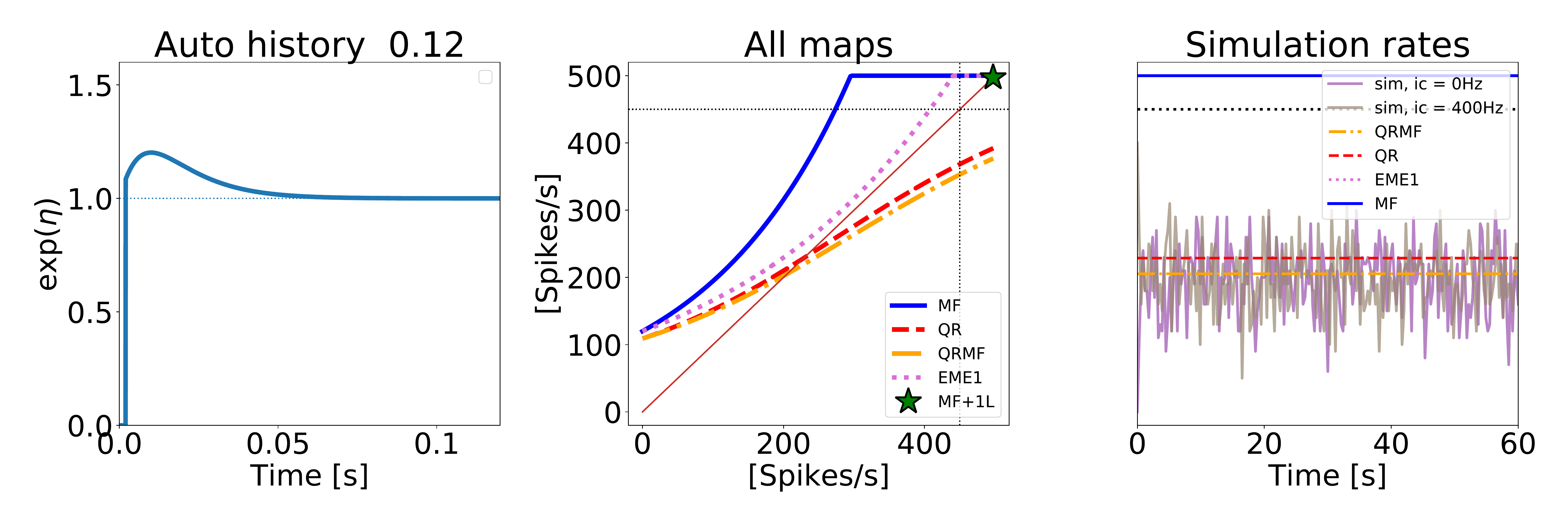}
    \caption{1D example: divergence predictions based on QR/QRMF and EME1/MF  maps can disagree. Auto-history filters were based on Erlang basis functions and an absolute refractory period.}
     \label{fig:MFworse}
  \end{figure}

\begin{figure}
      \vskip-1.5em
      \small
      \centering
      \includegraphics[width=0.45\textwidth]{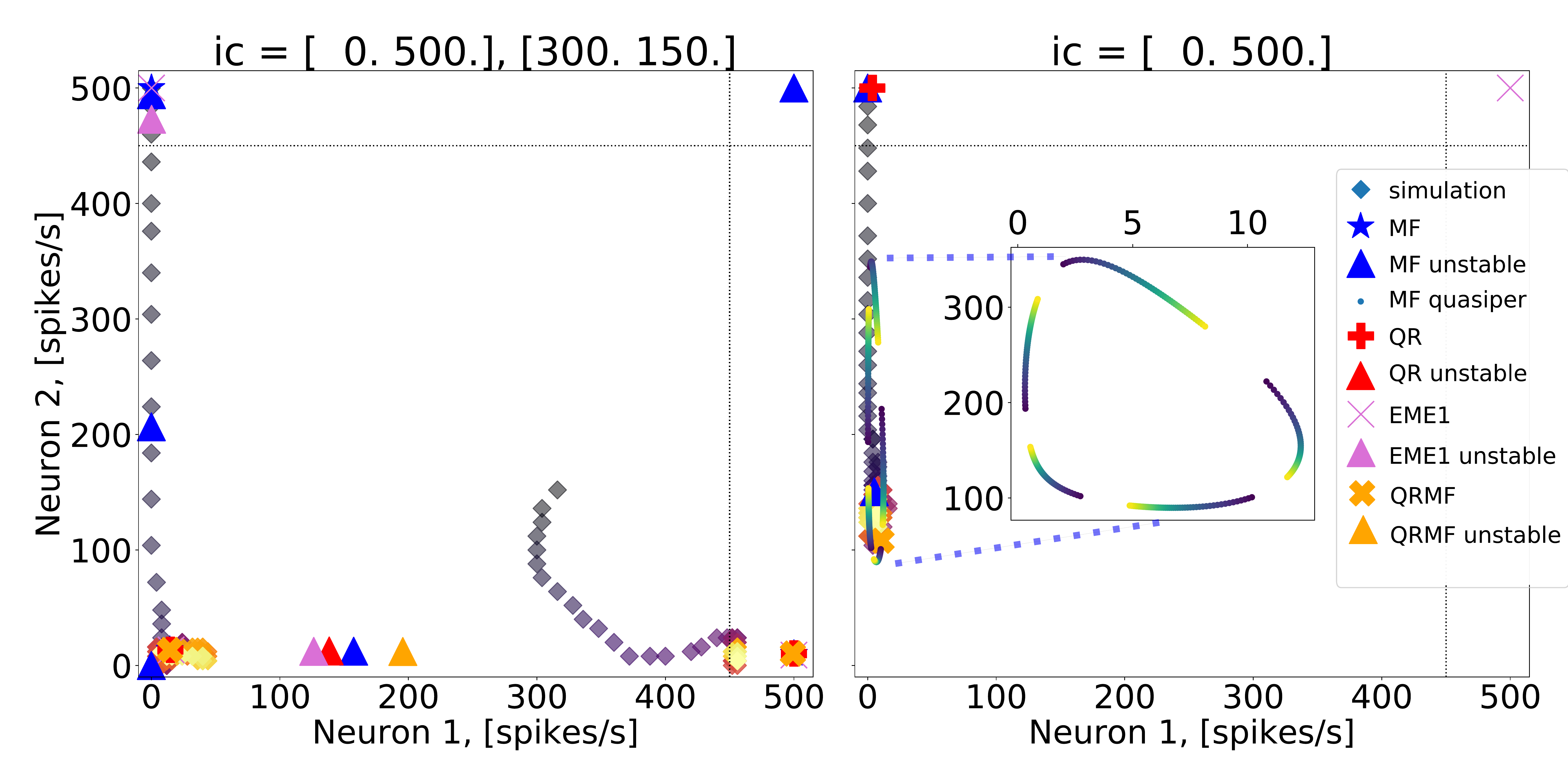}
      \caption{
      Theory and simulation of a 2D network. Multistability predicted by MF, QR, QRMF and EME1, and as observed in simulations. The MF approximation can show quasiperiodic dynamics (right panel and inset). ``ic'' denotes ``initial conditions'', and ``quasiper'' denotes '`quasiperiodic''.}
      \label{fig:2Dmultistab}
    \end{figure}

We note that the derived maps for the NHP may have multiple stable fixed points for stationary mean rates (i.e. multistability or metastability). MF/QR/EME1/QRMF maps can detect this multistability in some cases, but not always. In addition, these approximations can also  incorrectly indicate multistability in some cases (Figs. \ref{fig:QRworse}-\ref{fig:2Dmultistab}).
   
The exploration of the dynamics corresponding to the maps from each approximation also reveal not only stable fixed points, but also unstable fixed points (by using Newton method-like root-finding algorithms).

\subsection{Systematic 1D}

Figure \ref{fig:stabrasters} shows how varying the baseline rate and
the scaling or coefficients for the basis functions of the auto-history filter affects the divergence to nonphysiological rates and prediction accuracy.

\newlength{\tablistwidth}
\setlength{\tablistwidth}{0.24\textwidth}
\setlength{\tablistwidth}{0.21\textwidth}

\begin{figure}
\vskip6pt
    \centering
  \begin{subfigure}{\tablistwidth}
    \centering
    \includegraphics[width=\tablistwidth]{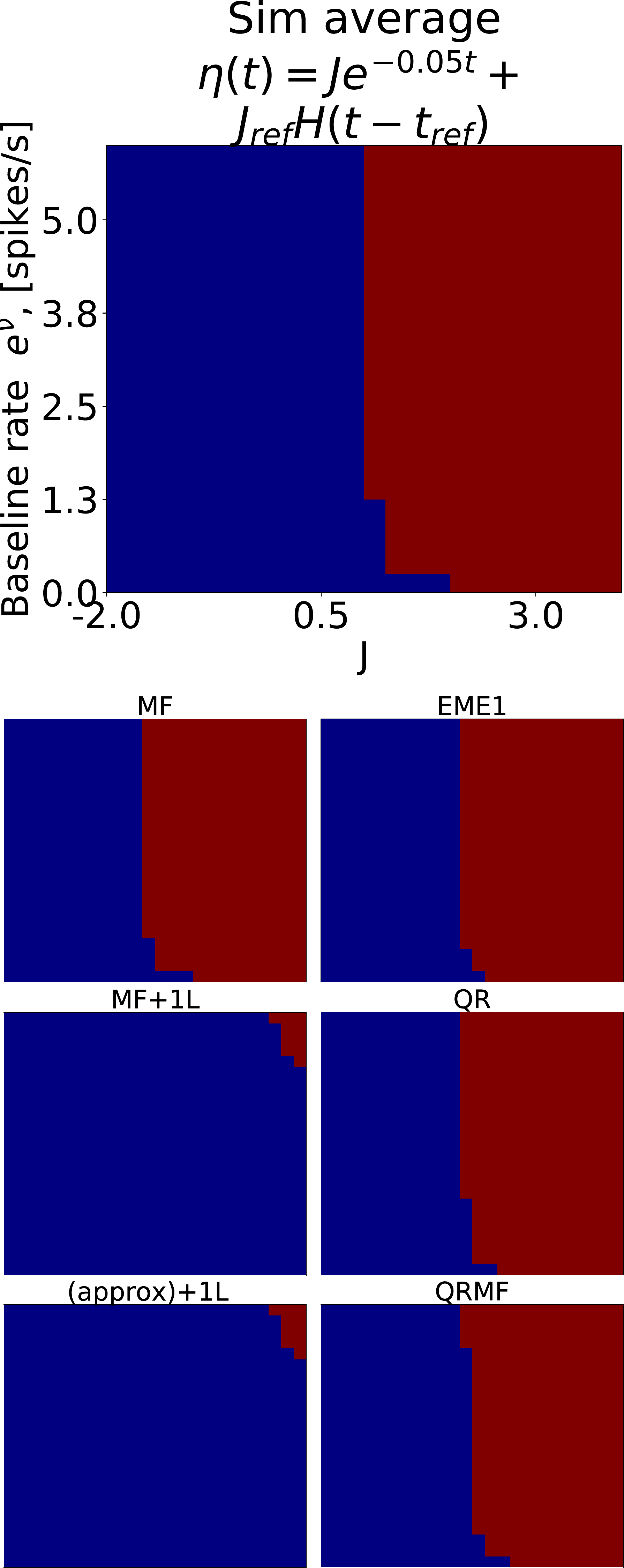}
  \end{subfigure}
    \hskip10pt
  \begin{subfigure}{\tablistwidth}
    \centering
    \includegraphics[width=\tablistwidth]{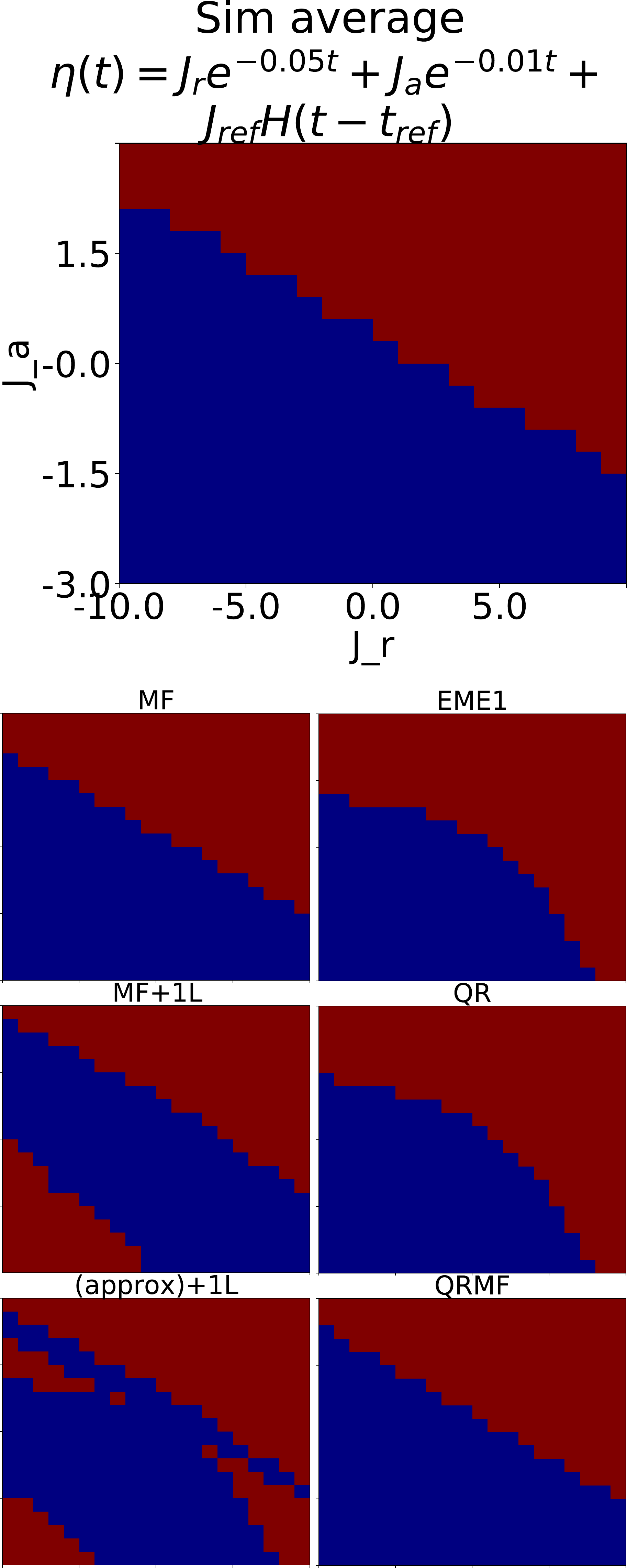}
  \end{subfigure}
  \caption{Stability plots. The two top show stability or divergence to nonphysiological rates as observed in simulations, depending the NHP model parameters (baseline and/or basis filter scale). Red indicates parameter regions where divergence happens, while blue indicates stability. The bottom plots show the same but now based on the different theoretical approximations.  }
  \label{fig:stabrasters}
\end{figure}

\subsection{Assessment of theoretical predictions}

In this section, we assess and summarize the performance of the different theoretical approximations applied to NHP models of dimension 1, 2 and 100.

%



\paragraph{D=1}

We have shown above how different theoretical approximation behave when the
filter shape is varied systematically in the parameter space. However, when
fitting PPGLMs to potentially more complex data from actual recordings, one
typically approximate auto- and cross-history filters with a larger number
(e.g. 10) of basis functions. Here, we attempted to systematically sample the
parameter space for the case of temporal filters constructed  or approximated
with 10 Erlang basis functions. In addition, we restricted these randomly
sampled filters to filters whose absolute values (beyond the refractory window)
and integrals are not very large  (threshold for absolute values = $20$,
threshold for absolute value of the intergral = $ \varkappa / \Delta t $)
and that they $\eta(\varkappa) < 0.1$ (we set the length $\varkappa$ of auto- and
cross-history filters to $200$ ms). To get the range of the 10-dimensional
parameters for the 10 basis functions, we first fitted the corresponding NHP
models ($\Delta t = 2$ ms) to a dataset containing neuronal ensemble recordings
obtained from monkey intracortical microelectrode array recordings during reach
and grasp actions \cite{gerhard2017stability}. The mean and $\pm 1$ standard
deviation of the fitted parameters were then used for sampling 10-dimensional
parameter space for auto-history filters. The parameters corresponding to the
baseline rates were generated in the same way.


\paragraph{D=2, two-exponential basis functions} We generated 2-neuron networks by randomly sampling auto- and cross-history filters consisting of the sum of two exponentials 
\begin{gather*}
  \eta_{\toind \toind}(t) =  J_{ref}H(t-\tau_{r} ) + J_r e^{-0.05t} + J_a e^{-0.01t}  \\
  \eta_{\toind\fromind}(t) =   J_r e^{-0.05t} + J_a e^{-0.01t} , \quad \fromind\neq \toind
\end{gather*}
where $J_{ref} = -10^{-6}$ captures the absolute refractory period for the auto-history filters, and $J_r$ and $J_a$ are the corresponding basis coefficients. 


Parameters for the auto-history filters were randomly (uniform) sampled from a box
centered at $(-6,-2)\times (-1,1) \subset \Rb^2$. Parameters for cross-history filters were taken
sampled from a box centered at $(-3.5,2.5)\times (-2,1) \subset  \Rb^2$. 


In addition, we randomly sampled (uniform) the log baseline rate from an interval centered at $-4.39$ with a SD $0.06$.
These parameter ranges were chosen in such a way that a considerable fraction
of the models diverged, so that both sensitivity and specificity of the
predictions of different theoretical approximations could be assessed. 

\paragraph{D=2, 10-dimensional (Erlang) basis functions} We also examined 2-neuron networks with temporal filters constructed from 10 (Erlang) basis functions. The history length was truncated at 200 ms. To systematically randomly sample different parameters for both auto- and cross-history filters, we first fitted 2D models to the same monkey recordings as above. In this case, however, we set the log baseline mean rate to $-2$ and set the standard deviation to $1.5$. As before, we discarded sampled filters that did not satisfy the 3 conditions specified above. 


\paragraph{D=100} We first fitted NHP to 100-neuron networks subsampled from the simulated Potjans-Diesmann model of cortical microcircuits \cite{potjans2012cell}. The model consists of 78,000 LIF neurons. Here, we increased the thalamic drive to layer 5 pyramidal neurons by 1.7 times and simulated the model for 200s. Next, we randomly subsample 100 subnetworks each of size 100 from the layer 5 pyramidal population. Finally, we fitted multivariate PPGLM models (10 Erlang basis functions, history length
truncated at 200 ms). WE note that the thalamic drive was increased with respect to typical values reported in the original publication in order to ensure that some of the estimated 100D NHP models diverged in numerical simulations. (None of the 100D NHP models fitted to subsampled networks from the Potjans-Diesmann model with thalamic
drive values from the original publication  diverged.)

 
Table \ref{table} summarizes the sensitivity and specificity to true divergence (as seen is simulation; see Methods) of the predictions based on all of the theoretical approximations, different dimensions and networks. 

\definecolor{lc}{rgb}{0.88,1,1}
\definecolor{mygray}{rgb}{0.9,0.9,0.9}

\addtolength{\tabcolsep}{-3pt}
 	
\newcolumntype{g}{>{\columncolor{mygray}}c}

  \begin{table}
  \vskip6pt
    \begin{tabular}{l|cg|cg|cg|cg}
      \textbf{Approx   / Dim} 
 &\multicolumn{2}{c}{Dim=    1} &\multicolumn{2}{c}{Dim=    2} &\multicolumn{2}{c}{Dim=    2b10} &\multicolumn{2}{c}{Dim=  100}
\\ pred quality                &  Sens      &   Spec     &  Sens      &   Spec     &  Sens      &   Spec     &  Sens      &   Spec    
\\ \hline
\\ MF                         &      100\% &       81\% &       72\% &       97\% &       90\% &       71\% &       88\% &       19\% 
\\ MF + 1L                &       21\% &       96\% &       99\% &       20\% &       27\% &       93\% &      100\% &        1\% 
\\ EME1                       &      100\% &       72\% &       80\% &       80\% &       91\% &       58\% &        0\% &      100\% 
\\ QR                         &      100\% &       83\% &       86\% &       90\% &       96\% &       76\% &      100\% &       13\% 
\\ QRMF                       &       99\% &       96\% &       66\% &      100\% &       93\% &       98\% &      100\% &       81\%
\\
\hline
\\
\hline
& \textbf{Ndiv} & \textbf{Ntot}   
& \textbf{Ndiv} & \textbf{Ntot}   
& \textbf{Ndiv} & \textbf{Ntot}   
& \textbf{Ndiv} & \textbf{Ntot}   
\\
 simulation                   &  301&1024  &      71&512 &     270&512  &       17&100  
    \end{tabular}                 
    \caption{Sensitivity and specificity for different dimensions, networks and theoretical approximations. The term Ndiv corresponds to the number of simulations that diverged to non-physiological rates out of a Ntot simulated networks. The case ``Dim=2 b10'' corresponds to 2-neuron networks whose auto- and cross-history filters were constructed/approximated with 10 basis functions.}
    \label{table}
\end{table}

\section{Discussion}
In this study we have examined the stability and dynamics of nonlinear Hawkes processes. In contrast to \cite{ocker2017linking}, here we have included the effects of absolute refractory periods and other auto-history effects. In contrast to \cite{naud2012coding,gerhard2017stability,chen2018stability}, we have extended the quasi-renewal (both QR and QRMF) approximations to the multivariate NHP case, i.e. neuronal networks. Also, in comparison to \cite{schwalger2017}, we used PPGLMs instead of GIF (no reset mechanism) and a discrete-time, instead of continuous, approximation. We note that MF multistability and its relation to runaway excitation has been studied in a different setting in \cite{schuecker2017fundamental}.

The three best approximations QRMF, QR and MF still show room for improvement. We conjecture \emph{diminished specificity} to arise from insufficient sensitivity to negative parts of the auto-history filter (QR case) and to the need for a better approximation of cross-history network effects (QR and QRFM). For lower-dimensions (D=1, D=2), we conjecture that the maximum of (non refractory part of) auto-history filter integral values determines the prediction outcome for MF,QR and QRMF approximations. MF+1L predicts instability for filters with large $\max_i( | \int \eta_{i\to i}| )$, i.e. for filters with \emph{large negative} integrals as well.

Looking at their definitions alone, we can contrast MF and QR approximations in terms of (1) the ratio of the contribution of $\pm$ phases (sign) of the temporal filters and of (2) inhomogeneities of spike process. QR is worse than MF regarding (1), but better on (2) since non-constant filter shapes do introduce variability. 
%
Nevertheless, we also note that while QR puts more weight on spike arrival times,
it still misses bursting effects \cite{naud2012coding}.

We described how the existence of stable fixed points in the nonphysiological
mean rate region predicts divergence according to different theoretical
approximations. We were also able to detect periodic and aperiodic dynamics in
some of the maps derived from the theoretical approximations. The relation
between existence of unstable points, periodic and aperiodic trajectories
remains to be investigated thoroughly, but so far our analysis showed that
taking the existence of periodicity into consideration improves sensitivity,
but lowers specificity.  

In this study we concentrated on the exponential nonlinearity case and on
Hawkes processes with enforced absolute refractory periods. However, MF, MF+1L
and QRMF (but not EME1 and QR) can be used for other nonlinearities and non-refractory processes as well, which might be beneficial for other applications outside
Neuroscience.

\section{Summary and outlook}

\newcommand{\imply}{$\Rightarrow$\ }
We have assessed the sensitivity and specificity of predicted divergence to nonphysiological mean firing rates as derived from various theoretical approximations to stationary rates of nonlinear Hawkes process models applied to spiking neuronal networks. Although some of the approximations showed perfect sensitivity, there is room for improvement of their specificity. We hope to address this problem, as well to further the understanding of the qualitative dynamics of multivariate NHPs, in future studies. In addition, we hope to examine how ensemble subsampling (e.g. subsampling of the Potjans-Diesmann model) affects the different theoretical approximations and their predictions about stability.

\section*{ACKNOWLEDGMENT}
We thank the Center for Computation and Visualization at Brown University for their computational resources.


\bibliographystyle{IEEEtran}
\bibliography{neuroAllArticles.bib}

\begin{thebibliography}{10}
\providecommand{\url}[1]{#1}
\csname url@samestyle\endcsname
\providecommand{\newblock}{\relax}
\providecommand{\bibinfo}[2]{#2}
\providecommand{\BIBentrySTDinterwordspacing}{\spaceskip=0pt\relax}
\providecommand{\BIBentryALTinterwordstretchfactor}{4}
\providecommand{\BIBentryALTinterwordspacing}{\spaceskip=\fontdimen2\font plus
\BIBentryALTinterwordstretchfactor\fontdimen3\font minus
  \fontdimen4\font\relax}
\providecommand{\BIBforeignlanguage}[2]{{%
\expandafter\ifx\csname l@#1\endcsname\relax
\typeout{** WARNING: IEEEtran.bst: No hyphenation pattern has been}%
\typeout{** loaded for the language `#1'. Using the pattern for}%
\typeout{** the default language instead.}%
\else
\language=\csname l@#1\endcsname
\fi
#2}}
\providecommand{\BIBdecl}{\relax}
\BIBdecl

\bibitem{truccolo2016point}
W.~Truccolo, ``From point process observations to collective neural dynamics:
  Nonlinear hawkes process glms, low-dimensional dynamics and coarse
  graining,'' \emph{Journal of Physiology-Paris}, vol. 110, no.~4, pp.
  336--347, 2016.

\bibitem{pillow2008spatio}
J.~W. Pillow, J.~Shlens, L.~Paninski, A.~Sher, A.~M. Litke, E.~Chichilnisky,
  and E.~P. Simoncelli, ``Spatio-temporal correlations and visual signalling in
  a complete neuronal population,'' \emph{Nature}, vol. 454, no. 7207, p. 995,
  2008.

\bibitem{Truccolo2005}
W.~Truccolo, U.~T. Eden, M.~R. Fellows, J.~P. Donoghue, and E.~N. Brown, ``{{A}
  point process framework for relating neural spiking activity to spiking
  history, neural ensemble, and extrinsic covariate effects},'' \emph{J.
  Neurophysiol.}, vol.~93, no.~2, pp. 1074--1089, Feb 2005.

\bibitem{truccolo2010collective}
W.~Truccolo, L.~R. Hochberg, and J.~P. Donoghue, ``Collective dynamics in human
  and monkey sensorimotor cortex: predicting single neuron spikes,''
  \emph{Nature neuroscience}, vol.~13, no.~1, p. 105, 2010.

\bibitem{weber2017capturing}
A.~I. Weber and J.~W. Pillow, ``Capturing the dynamical repertoire of single
  neurons with generalized linear models,'' \emph{Neural computation}, vol.~29,
  no.~12, pp. 3260--3289, 2017.

\bibitem{gerhard2017stability}
F.~Gerhard, M.~Deger, and W.~Truccolo, ``On the stability and dynamics of
  stochastic spiking neuron models: Nonlinear hawkes process and point process
  glms,'' \emph{PLoS computational biology}, vol.~13, no.~2, p. e1005390, 2017.

\bibitem{chen2018stability}
Y.~Chen, Q.~Xin, V.~Ventura, and R.~E. Kass, ``Stability of point process
  spiking neuron models,'' \emph{Journal of computational neuroscience}, pp.
  1--14, 2018.

\bibitem{ocker2017linking}
G.~K. Ocker, K.~Josi{\'c}, E.~Shea-Brown, and M.~A. Buice, ``Linking structure
  and activity in nonlinear spiking networks,'' \emph{PLoS Computational
  Biology}, vol.~13, no.~6, p. e1005583, 2017.

\bibitem{potjans2012cell}
T.~C. Potjans and M.~Diesmann, ``The cell-type specific cortical microcircuit:
  relating structure and activity in a full-scale spiking network model,''
  \emph{Cerebral cortex}, vol.~24, no.~3, pp. 785--806, 2012.

\bibitem{gerstner2014neuronal}
W.~Gerstner, W.~M. Kistler, R.~Naud, and L.~Paninski, \emph{Neuronal dynamics:
  From single neurons to networks and models of cognition}.\hskip 1em plus
  0.5em minus 0.4em\relax Cambridge University Press, 2014.

\bibitem{pozzorini2015automated}
C.~Pozzorini, S.~Mensi, O.~Hagens, R.~Naud, C.~Koch, and W.~Gerstner,
  ``Automated high-throughput characterization of single neurons by means of
  simplified spiking models,'' \emph{PLoS computational biology}, vol.~11,
  no.~6, p. e1004275, 2015.

\bibitem{bremaud1996stability}
P.~Br{\'e}maud and L.~Massouli{\'e}, ``Stability of nonlinear hawkes
  processes,'' \emph{The Annals of Probability}, pp. 1563--1588, 1996.

\bibitem{daley2007introduction}
D.~J. Daley and D.~Vere-Jones, \emph{An introduction to the theory of point
  processes: volume II: general theory and structure}.\hskip 1em plus 0.5em
  minus 0.4em\relax Springer Science \& Business Media, 2007.

\bibitem{baccelli2013elements}
F.~Baccelli and P.~Br{\'e}maud, \emph{Elements of queueing theory: Palm
  Martingale calculus and stochastic recurrences}.\hskip 1em plus 0.5em minus
  0.4em\relax Springer Science \& Business Media, 2013, vol.~26.

\bibitem{naud2012coding}
R.~Naud and W.~Gerstner, ``Coding and decoding with adapting neurons: a
  population approach to the peri-stimulus time histogram,'' \emph{PLoS
  computational biology}, vol.~8, no.~10, p. e1002711, 2012.

\bibitem{van1992stochastic}
N.~G. Van~Kampen, \emph{Stochastic processes in physics and chemistry}.\hskip
  1em plus 0.5em minus 0.4em\relax Elsevier, 1992, vol.~1.

\bibitem{schwalger2017}
T.~Schwalger, M.~Deger, and W.~Gerstner, ``Towards a theory of cortical
  columns: From spiking neurons to interacting neural populations of finite
  size,'' \emph{PLoS computational biology}, vol.~13, no.~4, p. e1005507, 2017.

\bibitem{duarte2016stability}
A.~Duarte, E.~L{\"o}cherbach, and G.~Ost, ``Stability, convergence to
  equilibrium and simulation of non-linear hawkes processes with memory kernels
  given by the sum of erlang kernels,'' \emph{arXiv preprint arXiv:1610.03300},
  2016.

\bibitem{schuecker2017fundamental}
J.~Schuecker, M.~Schmidt, S.~J. van Albada, M.~Diesmann, and M.~Helias,
  ``Fundamental activity constraints lead to specific interpretations of the
  connectome,'' \emph{PLoS computational biology}, vol.~13, no.~2, p. e1005179,
  2017.

\end{thebibliography}




\vspace{.3in}

\section*{Errata: \\ Stability of stochastic finite-size spiking-neuron networks: Comparing mean-field, 1-loop correction and quasi-renewal approximations}

\vspace{.3in}

\flushleft \textbf{Abstract} This errata note corrects typos and clarifies aspects related to equations presented in \textit{Proc IEEE EMBC (2019), pp. 4380 - 4386,} doi 10.1109/EMBC.2019.8857101.

\vspace{.2in}


\flushleft The article included above corresponds to the corrected version of the original article \textbf{[1]}. This errata note refers to the pagination in \textbf{[1]}. These are the corrections:

\vspace{.15in}

\textbf{(A)} The correct equation for solving the ``linear response'', or
``tree-level'' response (page 4382), reads
\begin{gather*}
    \bar{\Delta}_{\toind\fromind}(t-t')  = \phi^{(1)}(\bar{r}_{\toind} )  
    \sum_k (\eta_{k \to i} * \bar{\Delta}_{kj} )(t-t')
    + \delta(t-t')\delta_{\toind\fromind},
\end{gather*}
where $\phi^{(1)}$ is the first-derivative of the nonlinearity.

Similarly,
\begin{gather*}
    \tilde\Delta_{\toind\fromind}(t-t')  = \phi^{(1)}(\bar{r}_{\toind} )  
    \sum_k (\tilde\eta_{k \to i} * \tilde\Delta_{kj} )(t-t')
    + \delta(t-t')\delta_{\toind\fromind}.
\end{gather*}


\textbf{(B)} The first step in the quasi-renewal (QR) approximation for the conditional intensity function of the nonlinear Hawkes process (page 4382) is
to formulate 
\[
		\begin{aligned}
		\lambda_i(t\vert N[-\infty,t)^d ) \approx & \exp \left( \nu_i + \eta_{i \to i}(t_i') \right) \times \\
		&\aver{\exp\left((\eta_{i \to i} * dN_i) (t_i')\right)}_{N_{[-\infty,t_i')}} \\
		&\times \exp \left(\sum_{j\neq i}^d \aver{(\eta_{j \to i} * dN_j) (t)} \right), 
		\end{aligned}
\]		
which leads to (Eq. 5, 4382) 
\begin{gather}
  \lambda_{\toind}^{QR}(s,A) :=  
  \exp \left( \nu_{\toind} + \eta_{\toind\to \toind}(s)+ 
  \vphantom{ A_i \int_{s}^{\infty} (e^{\eta_{\toind\to \toind}(z)}-1) \, dz   + \sum_{\toind\neq \fromind}^d A \int_{0}^{\infty} {\eta_{\fromind\to \toind} (z) } }
  \right.
  \label{eq:lambda0}
  \nonumber
  \\
  \left.
 + A_i \int_{s}^{\infty} (e^{\eta_{\toind\to \toind}(z)}-1) \, dz   
  + \sum_{\fromind\neq \toind}^d A_{\fromind} \int_{0}^{\infty} {\eta_{\fromind\to \toind} (z) }  \, dz   \right), 
  \nonumber
\end{gather}
where $A$ is a vector whose components are given by the average mean firing rates of the corresponding neurons (we consider a stationary case), and $s = t - t_{\toind}$ is the time elapsed since the most recent spike. 

\vspace{.15in}

\textbf{(C)} The quasi-renewal mean-field approximation (Eq. 7, page 4383) is
\begin{gather*}
  \lambda_{\toind}^{QRMF}(s,A) :=  
  \exp \left( 
  \vphantom{ 
  A \int_{s}^{\infty} \eta_{\toind\to \toind}(z)  dz  + 
  \sum_{\toind\neq \fromind} A \int_{0}^{\infty} {\eta_{\fromind\to \toind} (z) }  dz   }
  \nu_{\toind} + \eta_{\toind\to \toind}(s)+ \right.
 \\
 \left.
 +
   A_i \int_{s}^{\infty} \eta_{\toind\to \toind}(z)  dz  + 
  \sum_{\fromind \neq \toind} A_{\fromind} \int_{0}^{\infty} {\eta_{\fromind\to \toind} (z) }  dz   \right). 
  \label{eq:lambda0MF}
\end{gather*}


\textbf{(D)} Equation 8 (page 4383) is replaced by the following: Given an approximation of a stationary rate $\bar{q}$, the 1-loop correction to a 
(mean field-,QR-,QRMF-, or EME1-derived) stability matrix $\Psi$ is
\begin{gather*}
\Gamma_1=\frac{1}{2} \phi^{(2)} \left(\nu+\left(\int\tilde{\eta} \right) \bar{q} \right)
\int_{0}^t     
\int_{0}^t 
\left(  (\tilde{\eta} * {\tilde{\Delta} }^{(\bar{q})} )(t_1-t_2)\right)^2 \\
\phi^{(1)}\left(\nu+\left(\int\tilde{\eta} \right) \bar{q} \right) \tilde{\eta} \   dt_2 \, dt_1,
\end{gather*}
where $\phi^{(2)}$ is the second-derivative of the nonlinearity. The one-loop corrected stability matrix is then
\begin{gather*}
\Psi_1 = \Psi + \Gamma_1.
\end{gather*}

\textbf{Note:} The numerical results and analyses presented in \textbf{[1]} were based on the correct expressions and equations.

\section*{Reference}
\textbf{[1]} Todorov D, Truccolo W (2019) Stability of stochastic finite-size spiking-neuron networks: Comparing mean-field, 1-loop correction and quasi-renewal approximations. Proceedings of the 41st Annual International Conference of the IEEE Engineering in Medicine and Biology Society (EMBC), pp. 4380 - 4386. doi 10.1109/EMBC.2019.8857101

\end{document}